\documentclass[twocolumn]{aastex631}

\shorttitle{Imaging Venus-like Worlds}
\shortauthors{Stephen R. Kane et al.}

\begin{document}

\title{Imaging Venus-like Worlds: Spectral, Polarimetric, and UV Diagnostics for the Habitable Worlds Observatory}

\author[0000-0002-7084-0529]{Stephen R. Kane}
\affiliation{Department of Earth and Planetary Sciences, University of California, Riverside, CA 92521, USA}
\email{skane@ucr.edu}

\author[0000-0002-4420-0560]{Kimberly M. Bott}
\affiliation{SETI Institute, Mountain View, CA 94043, USA}

\author[0000-0002-4258-6703]{Kenneth E. Goodis Gordon}
\affiliation{Planetary Sciences Group, Department of Physics, University of Central Florida, 4111 Libra Drive, Orlando, FL 32816, USA}
\affiliation{Space Research Institute, Austrian Academy of Sciences, Schmiedlstrasse 6, A-8042 Graz, Austria}

\author[0009-0006-9233-1481]{Emma L. Miles}
\affiliation{Department of Earth and Planetary Sciences, University of California, Riverside, CA 92521, USA}

\author[0000-0001-7968-0309]{Colby M. Ostberg}
\affiliation{Department of Earth and Planetary Sciences, University of California, Riverside, CA 92521, USA}

\author[0000-0002-5644-7069]{Paul K. Byrne}
\affiliation{Department of Earth, Environmental, and Planetary Sciences, Washington University, St. Louis, MO 63130, USA}

\author[0000-0001-9355-3752]{Ludmila Carone}
\affiliation{Space Research Institute, Austrian Academy of Sciences, Schmiedlstrasse 6, A-8042 Graz, Austria}

\author[0000-0002-6939-9211]{Tansu Daylan}
\affiliation{Department of Physics and McDonnell Center for the Space Sciences, Washington University, St. Louis, MO 63130, USA}

\author[0000-0003-1756-4825]{Antonio Garc\'ia Mu\~noz}
\affiliation{Universit\'e Paris-Saclay, Universit\'e Paris Cit\'e, CEA, CNRS, AIM, Gif-sur-Yvette, 91191, France}

\author[0000-0001-5737-1687]{Caleb K. Harada}
\altaffiliation{NSF Graduate Research Fellow}
\affiliation{Department of Astronomy, 501 Campbell Hall \#3411, University of California, Berkeley, CA 94720, USA}

\author[0000-0003-2215-8485]{Renyu Hu}
\affiliation{Jet Propulsion Laboratory, California Institute of Technology, Pasadena, CA 91109, USA}
\affiliation{Department of Astronomy \& Astrophysics, The Pennsylvania State University, University Park, PA 16802, USA}
\affiliation{Center for Exoplanets and Habitable Worlds, The Pennsylvania State University, University Park, PA 16802, USA}
\affiliation{Institute for Computational and Data Science, The Pennsylvania State University, University Park, PA 16802, USA}

\author[0000-0003-1629-6478]{Noam. R. Izenberg}
\affiliation{Johns Hopkins University Applied Physics Laboratory, Laurel, MD 20723, USA}

\author[0000-0002-0000-199X]{Erika Kohler}
\affiliation{NASA Goddard Space Flight Center, 8800 Greenbelt Road, Greenbelt, MD 20771, USA}

\author[0000-0002-7670-670X]{Malena Rice}
\affiliation{Department of Astronomy, Yale University, 219 Prospect Street, New Haven, CT 06511, USA}

\author[0000-0002-6650-3829]{Sabina Sagynbayeva}
\affiliation{Department of Physics and Astronomy, Stony Brook University, Stony Brook, NY 11794, USA}

\author[0000-0002-8053-1041]{Manuel Scherf}
\affiliation{Space Research Institute, Austrian Academy of Sciences, Schmiedlstrasse 6, A-8042 Graz, Austria}

\author[0000-0002-2949-2163]{Edward W. Schwieterman}
\affiliation{Department of Earth and Planetary Sciences, University of California, Riverside, CA 92521, USA}

\author[0000-0002-8900-3667]{Peter Woitke}
\affiliation{Space Research Institute, Austrian Academy of Sciences, Schmiedlstrasse 6, A-8042 Graz, Austria}


\begin{abstract}

Understanding planetary habitability requires a comparative approach that explores the divergent evolutionary outcomes of Earth and Venus. The Habitable Worlds Observatory (HWO) will be uniquely positioned to conduct a statistical and physical census of terrestrial exoplanets spanning the Venus Zone (VZ) and the Habitable Zone (HZ), enabling the detection and atmospheric characterization of post-runaway greenhouse worlds (``exoVenuses''). We present an updated list of VZ exoplanets, which raises the number of known candidates to 370. We describe a science case and an observing strategy for VZ exoplanets that integrates precursor exoplanet detection data and stellar characterization with HWO direct imaging, spectroscopy across the UV/optical/IR, and spectropolarimetry. Our proposed framework emphasizes a pathway toward the diagnosis of sulfur chemistry (SO$_2$) and aerosol physics (H$_2$SO$_4$ clouds/hazes), planetary redox states (O$_2$/O$_3$ false positives from hydrogen loss), and cloud microphysics detection (rainbow polarization). We quantify implications for HWO requirements, including UV access to 0.2--0.4~$\mu$m, optical/NIR coverage to $\gtrsim$1.5~$\mu$m, inner working angle (IWA) reaching 0.3--1.5~AU around nearby Sun-like stars, and the SNR/resolution needed for key features. Finally, we outline a community-driven path to producing robust demographic inferences and target selection for optimizing HWO observations.

\end{abstract}

\keywords{astrobiology -- planetary systems -- planets and satellites: dynamical evolution and stability -- planets and satellites: individual (Venus)}


\section{Introduction}
\label{intro}

Comparative planetology is key to understanding why Earth and Venus, which are nearly twins in bulk properties (size, mass, and likely initial volatile inventories, diverged so profoundly in climate and surface environment \citep{nimmo1998,bullock2001,kane2019d,way2020,gillmann2022,kane2024b}. Venus is an archetype of runaway greenhouse conditions, with a dense, CO$_2$-dominated atmosphere, a surface pressure a factor of $\sim$90 than Earth, pervasive sulfuric-acid cloud decks, and a surface temperature hot enough to preclude stable liquid water \citep{ingersoll1969c,kasting1988c}. Whereas Earth's surface is continually reshaped by plate tectonics and sustained ocean-atmosphere exchange, Venus lacks clear evidence for modern plate tectonics and instead appears to have experienced a distinct volcanic and resurfacing history, with limited present-day pathways for long-term CO$_2$ sequestration \citep{solomon1992,nimmo1998,smrekar2010a}. The two planets also differ in key atmospheric processes and boundary conditions: Venus's extremely dry atmosphere, high albedo from global clouds, and low velocity near-surface winds under a super-rotating atmosphere \citep{donahue1982,kasting1983,lebonnois2010,kane2022b}. The exoplanet era now offers an empirical route to generalizing the divergence of Venus and Earth by measuring the demographics and atmospheric physics of terrestrial worlds that receive Venus-similar fluxes (a factor of $\sim$1.91 larger than Earth) and comparing them to potentially temperate planets. Such measurements will provide invaluable data with which to diagnose the atmospheric and surface evolution of these planets, thus testing the concepts of the Venus Zone (VZ) \citep{kane2014e,vidaurri2022b,ostberg2023a,miles2025} and the Habitable Zone (HZ) \citep{kasting1993a,kane2012a,kopparapu2013a,kopparapu2014,kane2016c,hill2018,hill2023}. Direct imaging and spectroscopy with the Habitable Worlds Observatory (HWO) can substantially contribute to this endeavor by measuring reflected-light spectra and polarization signals of nearby small planets at star--planet separations of $\sim$0.3--1.5~AU around FGK (and select early-M) host stars, with precursor transit, radial-velocity (RV), and stellar programs furnish radii, masses, architectures, and abundance priors closing the loop between interiors and atmospheres \citep{kane2014e,kopparapu2014,meadows2017,meadows2018c,sagynbayeva2025c}.

There are two main aspects of HWO capabilities that enable a deeper dive into the identification of possible Venus-like exoplanets (or ``exoVenus" for short). First, access to the near-UV (0.2--0.4~$\mu$m) enables direct information on atmospheric sulfur dioxide (SO$_2$) and ozone (O$_3$), which together probe active sulfur photochemistry, potential volcanic outgassing, and photochemical redox pathways that can generate oxygenated atmospheres abiotically \citep{domagalgoldman2014,wordsworth2014,arney2016,meadows2017,meadows2018c}. Second, the combination of optical/NIR reflectance spectra (0.4--1.5~$\mu$m) and linear spectropolarimetry characterizes the presence, composition, and microphysics of aerosols, particularly H$_2$SO$_4$ clouds and hazes that flatten the spectral continuum and provide phase-dependent polarization ``rainbows'' diagnostics of droplet refractive indices \citep{hansen1974c,bailey2007b,lincowski2018}. Together, these diagnostics help break aerosol/gas degeneracies that complicate retrievals in the presence of thick clouds, and they complement the mid-IR strengths of the James Webb Space Telescope (JWST), which has already revealed evidence of high-altitude SO$_2$ in hot planetary atmospheres at $\sim$4~$\mu$m \citep{tsai2023b}.

From a population perspective, there are already hundreds of known exoplanets smaller than $2R_\oplus$ that lie within the VZ of their host stars, and this number is expected to grow substantially in the coming years \citep{ostberg2019,ostberg2023a}. Establishing the frequency of planets in a post-runaway greenhouse state as a function of stellar type and system architecture (including the role of exterior gas giants) is therefore within reach as the inventory of exoplanets continues to increase \citep{way2020,kane2024b,kane2025a}. Furthermore, synergy with ongoing and planned Venus missions will provide Solar System ``ground truth'' measurements for sulfur chemistry, cloud microphysics, and surface–atmosphere exchange that can be incorporated into exoplanet retrieval priors and interpretation frameworks \citep{garvin2022,cascioli2021,widemann2023,kane2019d,kane2021d,seager2022a}.

This paper presents a detailed description of the HWO instrument requirements and potential science return from a program that focuses on the study of Venus and its exoplanet analogs. Section~\ref{science} explains the science motivation for the study of Venus analogs and the need for HWO observations. The observational requirements that maximize the exoVenus science yield are outlined in Section~\ref{obsreq}, including the key diagnostics, target sample, and observing strategy. Section~\ref{pathway} provides a description of the modeling pathway, potential yields, and the synergy with Venus missions that will be enabled by the HWO exoVenus program. Section~\ref{discussion} discusses additional components of the observations and preparatory work for consideration, and we provide concluding remarks in Section~\ref{conclusions}.


\section{Science Motivation}
\label{science}

A central goal in planetary science and astrobiology is to map the physical conditions that allow temperate surface environments to form and persist over Gyr timescales. Within the Solar System, Venus and Earth provide a natural laboratory to explore divergent climate evolution, despite similar bulk properties and initial conditions \citep{kane2019d}. The degree to which insolation alone, versus rotation, obliquity, magnetic environment, volatile inventory, and interior-atmosphere coupling, drove the pathway of Venus to its present runaway greenhouse remains debated \citep{hamano2013,way2016,way2020,turbet2021,constantinou2025}. A complementary path forward is to expand the sample via exoplanets: characterizing terrestrial planets spanning the VZ and HZ continuum and quantifying the prevalence and diversity of Venus-like states. Existing and ongoing surveys have revealed hundreds of candidate exoVenuses with $R_p \lesssim 2$~$R_\oplus$ at VZ insolations \citep[e.g.,][]{kane2014e,ostberg2023a}. This inventory will grow as candidates from the Transiting Exoplanet Survey Satellite \citep[TESS;][]{ricker2015} continue to be confirmed \citep{ostberg2019}, and the prospect of many more transiting terrestrial exoplanet discoveries via the PLAnetary Transits and Oscillations of stars (PLATO) mission \citep{rauer2025}. Furthermore, non-transiting VZ exoplanets will be supplied in increasing numbers as extreme-precision radial velocity (EPRV) programs mature \citep{fischer2016,morgan2021b,kane2024e}. HWO will be able to directly image and spectrally characterize nearby systems, enabling analysis of atmospheres, cloud properties, and energy budgets at orbital separations overlapping 0.3--1.5~AU around Sun-like hosts.

\begin{figure*}
    \centering
    \includegraphics[width=0.8\textwidth]{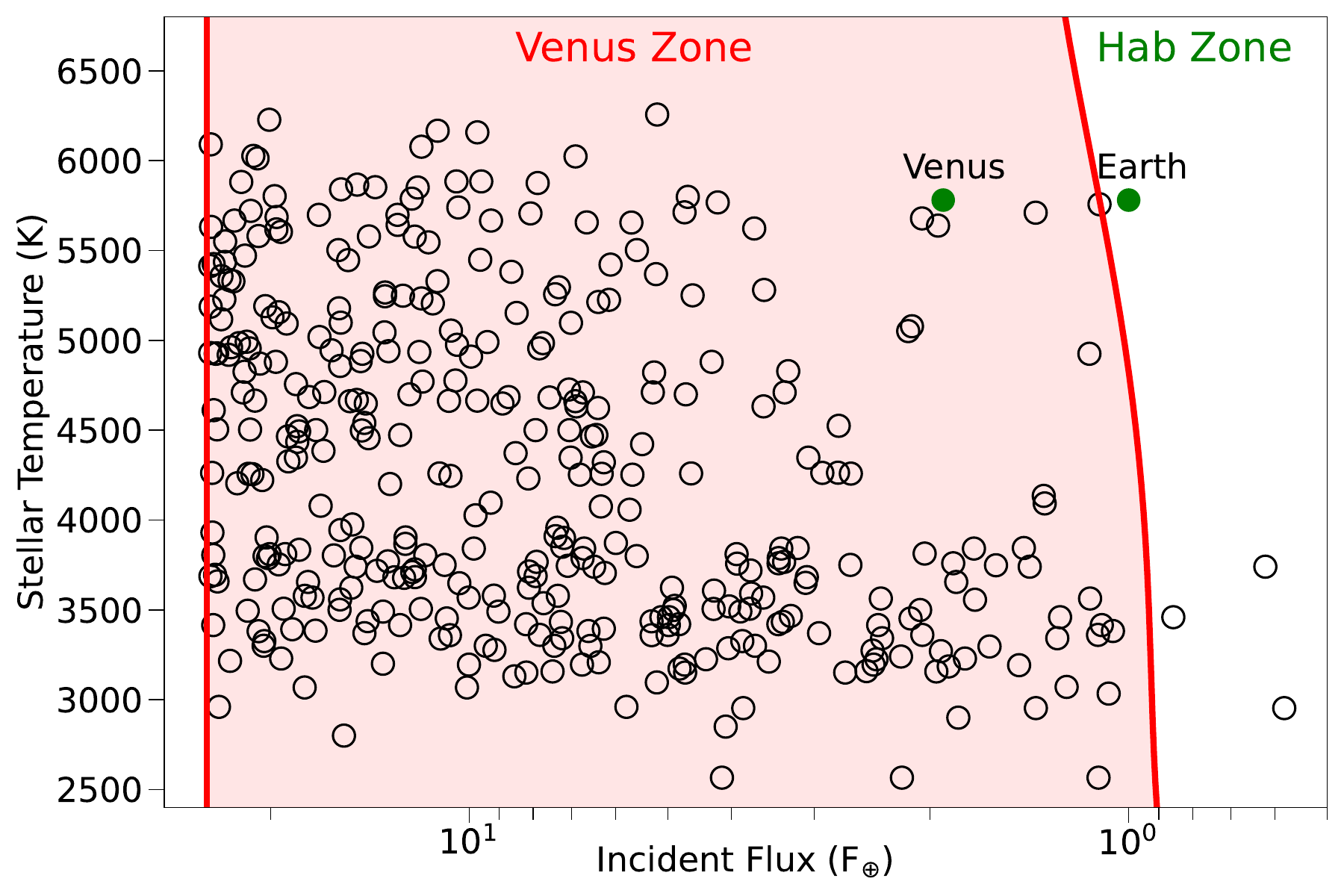}
    \caption{The distribution of currently known exoplanets from the NASA Exoplanet Archive that are in the VZ and have radii $R_p \lesssim 2$~$R_\oplus$. The inner VZ boundary is indicated by the red line on the left, and the outer VZ boundary is shown on the right. As of November 2025 there are 370 potential exoVenuses, and this number will continue to grow as there are over 7000 TESS planet candidates yet to be confirmed. The VZ will be calculated and translated against the IWA for all proposed HWO priority targets.}
    \label{fig:vz}
\end{figure*}

The overarching goal is to determine the occurrence, atmospheric composition, and climate states of exoVenus analogs, and to establish empirical connections to Solar System Venus \citep[e.g.,][]{kane2019d,way2020,kane2021d,kane2022b,wilson2022c,kane2024b}. In the near term, predictions for the HWO yields of terrestrial planets are underway, by merging current demographics with instrument performance, such as the inner working angle (IWA) and contrast ratio \citep{stark2024b}. Such efforts would optimally be supported by a coordinated precursor program that refines planetary masses, radii, and architectures, while characterizing host stars and activity \citep{harada2024b,kane2024e,harada2025}. Over the longer term, HWO observations will enable spectroscopic identification of post-runaway greenhouse states and retrieval of key atmospheric parameters, including SO$_2$ abundance, H$_2$SO$_4$ cloud/haze properties, continuum albedo, and proxies for thermal structure and redox. These ensemble results may be used to infer occurrence rates of Venus-like versus temperate states as a function of stellar type and system architecture \citep{kane2014e,kopparapu2014}. For example, shown in Figure~\ref{fig:vz} are the distribution of currently known confirmed terrestrial exoplanets (370 as of November 2025) that reside within the VZ \citep{ostberg2019,ostberg2023a}. Data were extracted from the NASA Exoplanet Archive \citep{akeson2013,christiansen2025}. The boundaries of the VZ are indicated by red lines, where the inner boundary is the complete atmospheric loss limit and the outer boundary is equivalent to the runaway greenhouse limit \citep{kane2014e}. The planets shown in Figure~\ref{fig:vz} all have radii $R_p < 2.0 $~$R_\oplus$, and spend any portion of their orbits within the boundaries of the VZ. Not included in this plot are planets that have null values for the host star effective temperature and planetary semi-major axis, or the means to calculate it. For planets with non-circular orbits, the shown incident flux is that received at the planet's semi-major axis. Planets without measured eccentricities were assumed to have circular orbits, which may have excluded a few planets in close proximity to either of the VZ boundaries \citep{ostberg2023a}.


\section{Observational Requirements}
\label{obsreq}


\subsection{Key Diagnostics for Venus Analogs}
\label{key}

Access to the near-UV (0.2--0.4~$\mu$m) is essential because strong SO$_2$ absorption provides a primary indicator of active sulfur chemistry and potential surface volcanism. Additionally, O$_3$ detection would help establish the atmospheric redox state and can break the degeneracy within the interpretation of oxygen detection for temperate planets \citep{arney2016,meadows2017,meadows2018c}. At visible wavelengths (0.4--1.0~$\mu$m), H$_2$SO$_4$ clouds and photochemical hazes produce a spectral slope that can suppress molecular bands, yielding flattened reflectance spectra reminiscent of Venus. The spectral continuum and phase dependence thus provide important information regarding cloud-top altitude, optical depth, and microphysics \citep{woitke2025}. Extending the observations into the NIR ($>$1.0--1.5~$\mu$m) improves sensitivity to particle size distributions and helps to disentangle aerosols from gas-phase opacity. For cooler hosts, coverage to $\sim$1.7--2.0~$\mu$m can add retrieval data on aerosol properties and, when present, surface/near-surface windows \citep{arney2018,lincowski2018}. Spectropolarimetry supplies complementary constraints to the above described data. Linear polarization typically peaks at optical wavelengths for Earth-like atmospheres, whereas Venus-like aerosol fields can shift and amplify signals into the NIR \citep{hansen1974c,garciamunoz2015a}. The phase angle and shape of the ``rainbow'' feature diagnose particle refractive index and thus droplet composition (e.g., water versus H$_2$SO$_4$), providing a composition test largely independent of spectroscopy \citep{goodisgordon2025a,roccetti2025b}. Phase curves in broadband and low-resolution modes (ideally coordinated with polarization cadence) further constrain measurements of albedo heterogeneity, cloud patchiness, and temporal variability while helping to locate the polarization rainbow and decouple cloud microphysics from surface contributions \citep[e.g.,][]{bailey2007b}.


\subsection{Target Sample and Demographics}
\label{sample}

An exoVenus target pool for HWO would ideally center on nearby FGK stars, with select early-M hosts, for which HWO can reach projected separations of 0.3--1.5~AU at the required contrasts \citep{ostberg2023a}. For planets that transit the host star, planetary radii may be extracted, which is a key component for deriving bulk density and surface/atmospheric albedo \citep{carriongonzalez2020}. For those systems with bright enough host stars, masses may be measured via either RV data or Transit Timing Variations (TTVs) exhibited within the photometry \citep{agol2005,holman2005,agol2021}, enabling atmospheric scale height estimates \citep{batalha2019c}. Atmospheric properties sampled from those planets amenable to transmission spectroscopy will provide valuable priors that contextualize HWO reflectance \citep{sagynbayeva2025c,tuchow2025a}. For non-transiting planets discovered by RVs, HWO imaging constrains estimates of inclination and thus the true planet mass, and reflectance spectroscopy or spectropolarimetry may then probe their atmospheres directly. As described in Section~\ref{science}, preliminary counts indicate several hundred sub-$2R_\oplus$ candidates in the VZ within reach of precursor characterization (see Figure~\ref{fig:vz}).

\begin{figure*}
    \begin{center}
        \begin{tabular}{cc}
            \includegraphics[width=8.5cm]{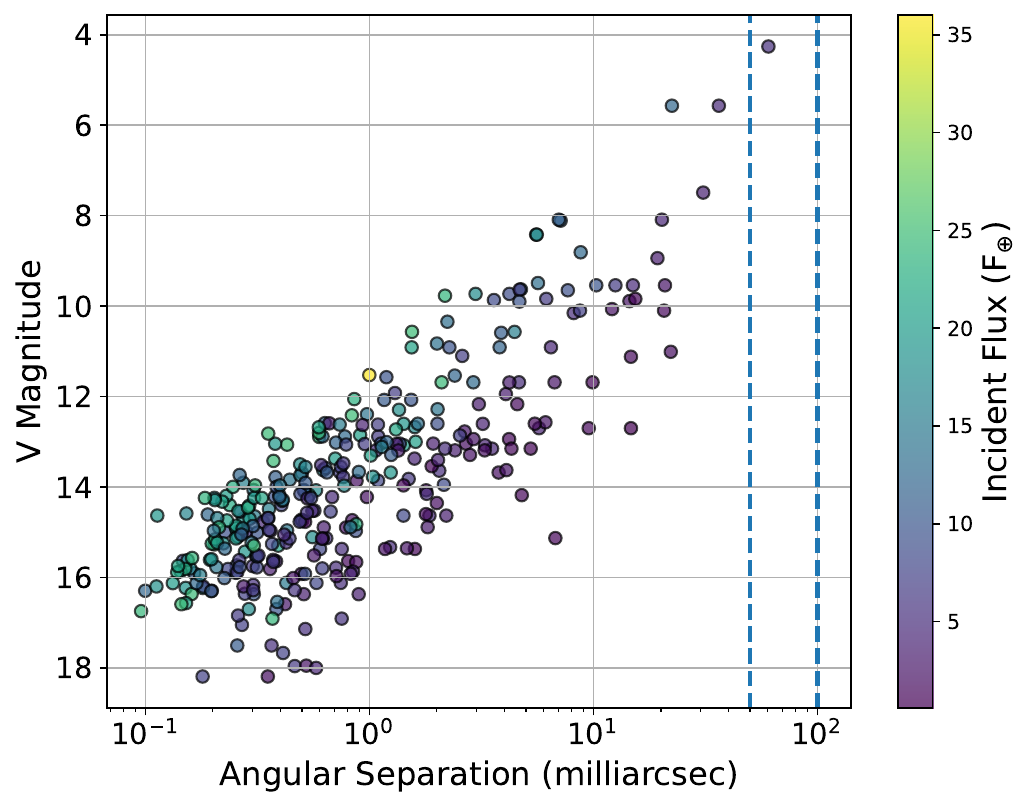} &
            \includegraphics[width=8.5cm]{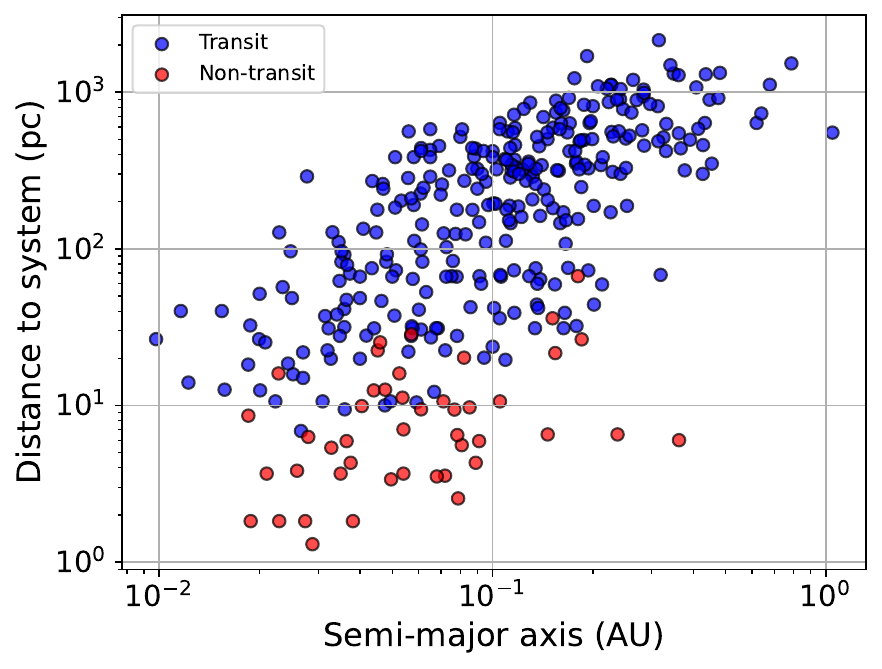} \\
        \end{tabular}
    \end{center}
  \caption{Properties of VZ candidates and their host stars. Left: Host star V magnitude, angular separation of the planet, and incident flux received by the planet. The two vertical dashed lines indicate an IWA of 50 and 100 mas. Right: Distance to the planetary systems and the semi-major axis of the planet. Systems where the planet is known to transit are shown in blue, otherwise they are shown in red.}
  \label{fig:targets}
\end{figure*}

However, due to the outer VZ boundary constraint and the observational bias toward shorter orbital periods within the transit and RV detection techniques \citep{borucki1984a,cumming2004,kane2005b,kane2007b,kane2008b}, VZ terrestrial planets tend to be relatively close to their host stars, presenting challenges to direct imaging prospects \citep{ostberg2023a}. Shown in Figure~\ref{fig:targets} are various relevant properties of VZ candidates and their host stars, using the same sample as described in Section~\ref{science} and shown in Figure~\ref{fig:vz}. The left panel shows the $V$ magnitude of the host star and the maximum angular separation of the planet ($\Delta \theta$) in milliarcseconds (mas), calculated from the semi-major axis ($a$) and system distance ($d$). The data points are color-coded by the incident flux on the planet ($F_p$), expressed in Earth units, spanning a wide range, from 0.58~$F_\oplus$ (Wolf 1061 d) to 36.01~$F_\oplus$ (Ross 176 b). Note that those cases with incident flux values less than unity are caused by eccentric orbits that enter the VZ, and represent interesting opportunities to test climate resilience to major changes in flux \citep{kane2012e,barnes2013a,way2017a,kane2020e,kane2021a}. Even so, the higher incident flux cases tend to correlate with smaller angular separation, which is as expected when averaged over system distance. Vast differences in incident flux will only influence the observed spectra insofar as they influence the signal-to-noise of the measurements. However, additional photochemistry may cause the atmosphere to deviate from a Venus atmosphere, thus resulting in a modified set of absorption features. The left panel of Figure~\ref{fig:targets} also shows two vertical dashed lines, which represent IWA values of 50~mas and 100~mas. Estimates for the HWO IWA vary, depending on the requirements of a particular science case, but generally vary in the range 40--80~mas at 0.5--1.0~$\mu$m \citep{vaughan2023,ware2025}. When ranking the VZ candidates by $\Delta \theta$, the top 20 candidates are shown in Table~\ref{tab:targets}, showing that HD 20794 d (a non-transiting planet) exceeds an angular separation of 50~mas.

Shown in the right panel of Figure~\ref{fig:targets} are the system distance and semi-major axis values for the VZ candidates, color coded by whether the planet transits (blue) or not (red). The data clearly show the bias of the transit method toward farther systems, which is necessary in order to provide enough stars within the survey to overcome the geometric transit probability \citep{borucki1984a}. Thus, the production of VZ candidate discoveries whose angular separation lies outside the HWO IWA requires relatively nearby systems and semi-major axes near the outer edge of the VZ. Although such discoveries may be produced during the survey phase of HWO, the majority will likely result from EPRV surveys that target the solar neighborhood stellar population with sub-meter per second precision that can access the terrestrial planet regime.

\begin{deluxetable}{lccccc} \label{tab:targets}
    \tablecaption{Sample of VZ exoplanet candidates.}
    \tablecolumns{6}
    \tablewidth{0pt}
    \tablehead{ 
        \colhead{Planet} & 
        \colhead{$a$ (AU)} &
        \colhead{$F_p$ ($F_\oplus$)} &
        \colhead{$V$ mag} &
        \colhead{$d$ (pc)} &        
        \colhead{$\Delta \theta$ (mas)}
        }
    \startdata 
HD 20794 d & 0.3625 & 5.21 & 4.26 & 6.00 & 60.389 \\
HD 219134 d & 0.2370 & 4.69 & 5.57 & 6.53 & 36.287 \\
GJ 411 b & 0.0788 & 3.75 & 7.49 & 2.55 & 30.945 \\
HD 219134 f & 0.1463 & 12.32 & 5.57 & 6.53 & 22.400 \\
Prox Cen d & 0.0288 & 1.81 & 11.01 & 1.30 & 22.141 \\
Barnard e & 0.0381 & 2.44 & 9.54 & 1.83 & 20.859 \\
Wolf 1061 c & 0.0890 & 1.28 & 10.10 & 4.31 & 20.669 \\
GJ 15 A b & 0.0720 & 4.25 & 8.09 & 3.56 & 20.212 \\
Gl 725 A b & 0.0680 & 3.35 & 8.94 & 3.52 & 19.311 \\
GJ 273 b & 0.0911 & 1.06 & 9.84 & 5.92 & 15.385 \\
Barnard c & 0.0274 & 4.71 & 9.54 & 1.83 & 15.001 \\
GJ 1061 d & 0.0540 & 0.58 & 12.70 & 3.67 & 14.703 \\
Ross 128 b & 0.0496 & 1.47 & 11.12 & 3.37 & 14.698 \\
GJ 251 b & 0.0808 & 2.37 & 9.89 & 5.58 & 14.479 \\
Barnard b & 0.0229 & 6.75 & 9.54 & 1.83 & 12.537 \\
GJ 625 b & 0.0784 & 2.07 & 10.07 & 6.47 & 12.107 \\
Barnard d & 0.0188 & 10.01 & 9.54 & 1.83 & 10.293 \\
L 98-59 f & 0.1052 & 1.10 & 11.69 & 10.62 & 9.906 \\
GJ 1061 c & 0.0350 & 1.38 & 12.70 & 3.67 & 9.530 \\
AU Mic d & 0.0853 & 12.55 & 8.81 & 9.72 & 8.771 \\
    \enddata
\tablenotetext{}{The full data set is available online.}
\end{deluxetable}


\subsection{Observing Strategy}
\label{strategy}

As described above, an intensive precursor effort will be required to complete inner architectures and mass measurements using EPRV facilities prior to HWO operations. The necessary measurements can be produced with instruments such as ESPRESSO, EXPRES, HARPS, KPF, NEID, and MAROON-X \citep{pepe2000,fischer2016,gibson2016,jurgenson2016b,schwab2016,seifahrt2016b,pepe2021}. Transit timing and duration measurements will refine masses and orbital dynamics where applicable, whereas stellar characterization campaigns will establish abundances (e.g., C/O, Mg/Si) and activity cycles that inform interior and atmospheric priors. For the relatively few targets that transit, a parallel JWST triage program will acquire transmission or emission spectra for the brightest exoVenus candidates to benchmark haze impacts and to define where HWO's UV/optical capabilities are most decisive. During HWO operations, we envision a direct-imaging campaign targeting at least $\sim$30 nearby systems sampling separations between 0.3 and 1.5~AU, with depths sufficient to detect Venus/Earth-size contrasts around Sun-like hosts. Spectral coverage will, when feasible, be contemporaneous across UV (0.2--0.4~$\mu$m), optical (0.4--1.0~$\mu$m), and NIR ($>$1.0--1.5~$\mu$m) to mitigate variability and to break aerosol/gas degeneracies \citep{salvador2024}. If HWO is equipped with a polarimetric instrument, spectropolarimetric observations in the optical and NIR will be scheduled at phase angles that maximize the rainbow diagnostic, and phase-resolved reflectance and polarization monitoring will map cloud microphysics, temporal variability, and atmospheric heterogeneities. Temporal variability may indicate the detection of surface features or, alternatively, atmospheric perturbations in the form of waves \citep{lee2020b}. Table~\ref{tab:reqs} summarizes how the science priorities of an exoVenus program similar to that described translate into instrument and mission capabilities.

\begin{deluxetable*}{l c c l}
\tablecaption{Science-driven capability guidance for HWO.\label{tab:reqs}}
\tabletypesize{\footnotesize}
\tablewidth{0pt}
\tablehead{
\colhead{Capability / Requirement} & \colhead{Necessary} & \colhead{Desired} & \colhead{Justification / Comments}
}
\startdata
UV access (0.2--0.4~$\mu$m) & Yes & Yes & \parbox[t]{0.55\textwidth}{SO$_2$ and O$_3$ leverage; redox context; breaks haze/gas degeneracies.} \\
Optical (0.4--1.0~$\mu$m) & Yes & Yes & \parbox[t]{0.55\textwidth}{H$_2$SO$_4$ haze slopes; phase curves for clouds/albedo.} \\
NIR coverage ($>$1.5~$\mu$m) & \nodata & Yes & \parbox[t]{0.55\textwidth}{Added leverage on aerosol size distribution and continuum constraints.} \\
Contemporaneous UV+opt+NIR & No & Yes & \parbox[t]{0.55\textwidth}{Mitigates variability; improves joint retrievals.} \\
High spatial resolution (small IWA) & Yes & Yes & \parbox[t]{0.55\textwidth}{Access to 0.3--1.5~AU separations for nearby FGK systems.} \\
Spectral resolving power & Yes & Yes & \parbox[t]{0.55\textwidth}{Adequate $R$ to separate continua vs.\ bandheads and polarization features.} \\
Spectropolarimetry & \nodata & Yes & \parbox[t]{0.55\textwidth}{Rainbow phase for droplet refractive index (water vs.\ H$_2$SO$_4$).} \\
Field of view / regard & No & Yes & \parbox[t]{0.55\textwidth}{Efficient scheduling; seasonal visibility of multiple targets.} \\
Rapid response & \nodata & \nodata & \parbox[t]{0.55\textwidth}{Not a primary driver; cadence planning more important.} \\
Phase-resolved scheduling & No & Yes & \parbox[t]{0.55\textwidth}{Map reflectance curves and polarization rainbows.} \\
\enddata
\tablecomments{\nodata\ denotes ``not required'' or ``not applicable'' for that column.}
\end{deluxetable*}


\section{Pathway to ExoVenus Characterization}
\label{pathway}


\subsection{Retrievals and Modeling}

Concise data interpretation requires using end-to-end retrieval simulations that couple forward models and Bayesian inversion to define the spectral resolution, wavelength ranges, and SNR necessary to detect expected signatures at $\geq 5\sigma$ across host spectral types and planetary parameters. Radiative-transfer calculations, for instance with the Planetary Spectrum Generator (PSG) \citep{villanueva2018b}, have been configured for Venus-like atmospheres with $\sim$10~bar surface pressures, SO$_2$ absorption in the UV, and H$_2$SO$_4$ clouds that mute optical features and flatten the continuum. We paid particular attention to degeneracies between aerosol slope and gas bands and to the role of UV coverage in mitigating these degeneracies. Stellar abundance constraints feed interior models to bound volatile budgets and hence plausible atmospheric compositions. It is also important to explore redox pathways that yield O$_2$/O$_3$ without biology through hydrogen escape on hot, hazy planets, thereby strengthening biosignature interpretation frameworks for temperate targets observed in parallel HWO programs.

\begin{figure*}
    \begin{center}
        \begin{tabular}{cc}
            \includegraphics[width=8.5cm]{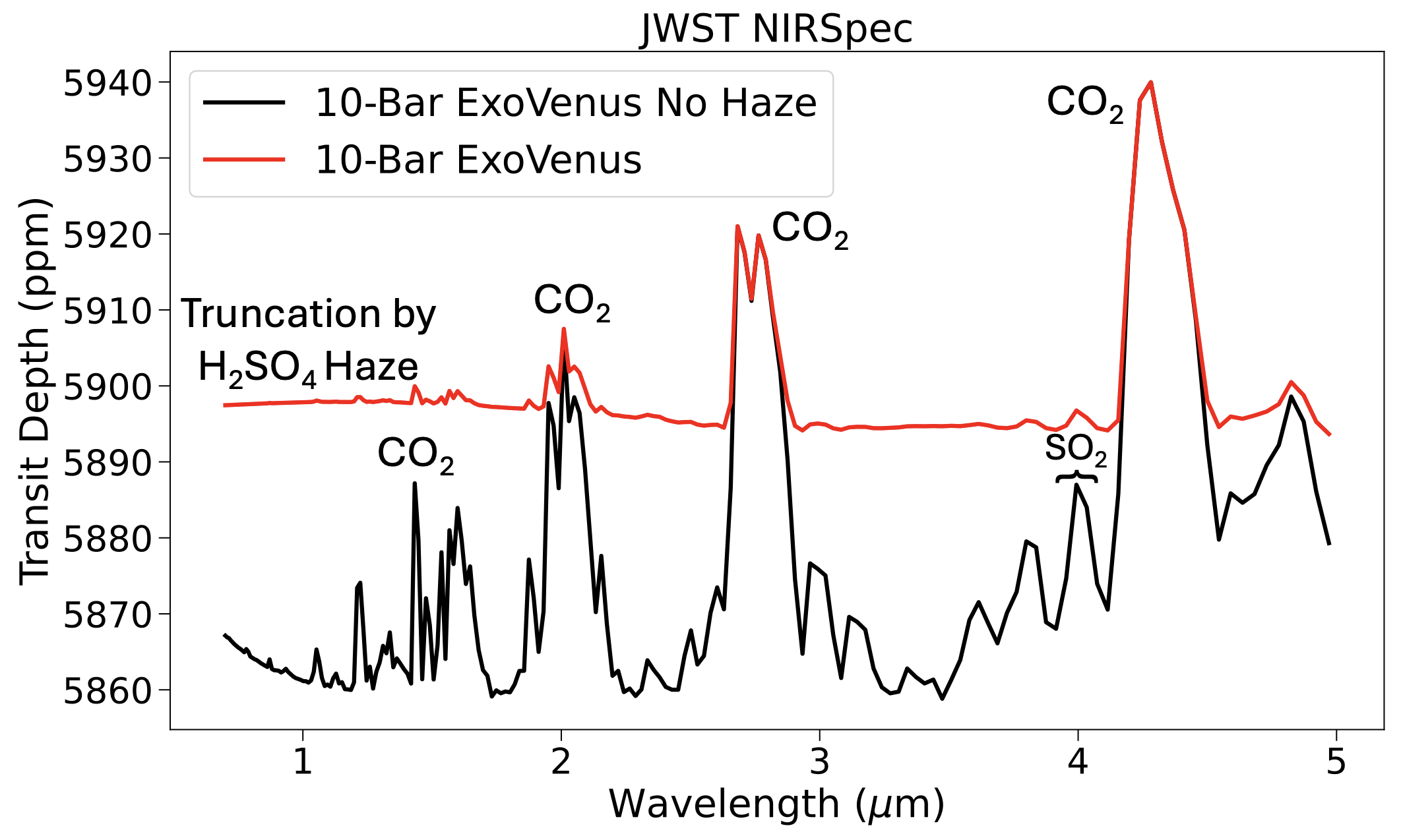} &
            \includegraphics[width=8.5cm]{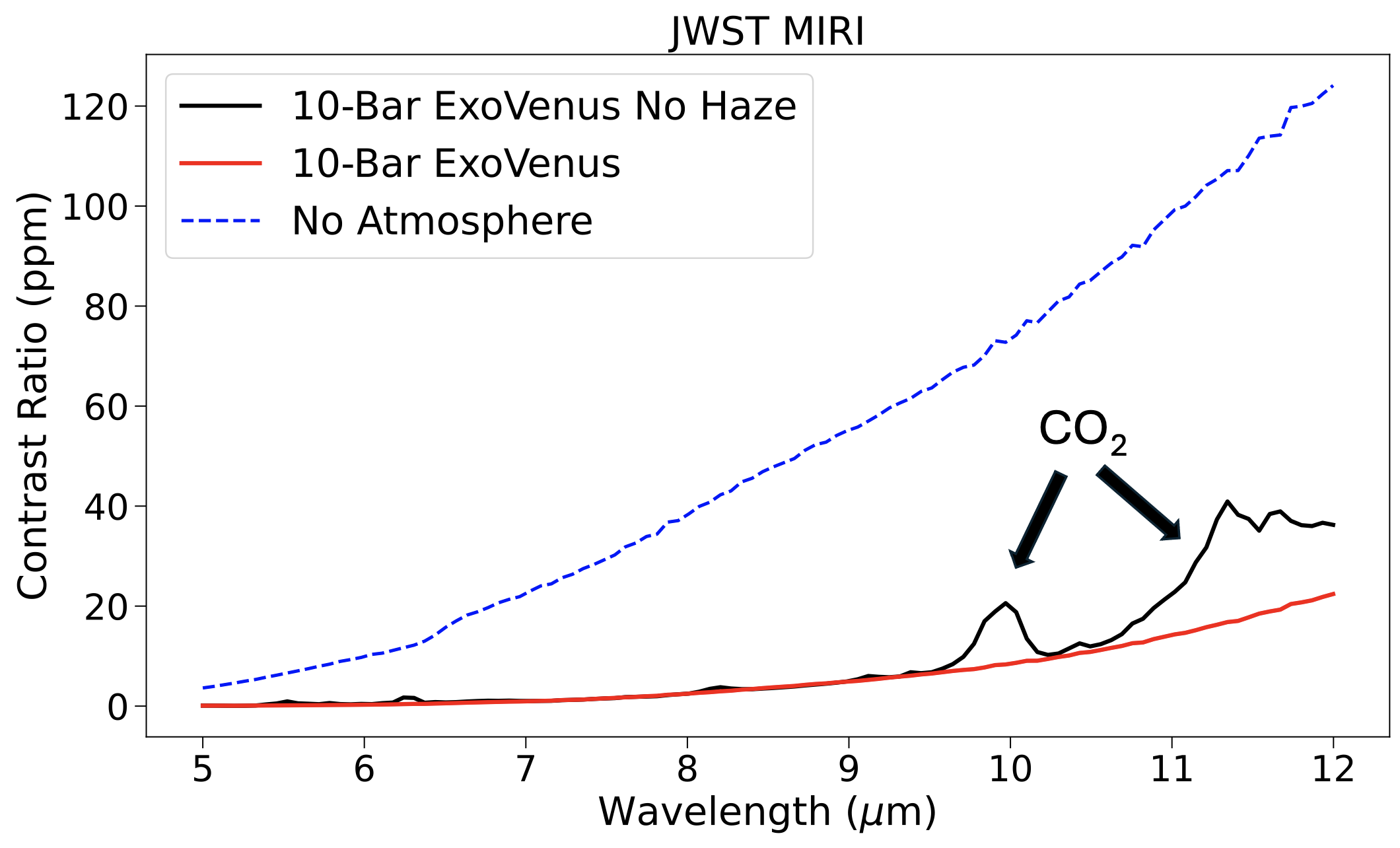} \\
        \end{tabular}
    \end{center}
   \vspace{-0.6cm}
  \caption{Modeled JWST observations of an exoplanet with a 10-bar, Venus-like atmosphere orbiting an M-type star. Left: A modeled transit spectrum of the exoplanet, with and without haze, if observed using JWST NIRSpec PRISM. The opaque haze layer truncates much of the spectrum, making CO$_2$ the only potentially detectable absorption feature. Right: A modeled eclipse spectrum of the exoplanet using JWST MIRI. The presence of an atmosphere may be able to be deduced from this spectrum, although the haze would inhibit estimates of the atmosphere's composition.}
  \label{fig:jwstspectra}
\end{figure*}

For example, we simulated the expected spectra for an exoplanet with a 10-bar, Venus-like atmosphere (96.5\% CO$_2$, 3.5\% plus additional trace gases as per the PSG Venus atmospheric composition template) using PSG \citep{villanueva2018b}. Figure~\ref{fig:jwstspectra} shows modeled transit (left panel) and eclipse (right panel) spectroscopy of the hypothetical exoVenus orbiting an M-type star at a separation of equivalent Venus flux if it were to be observed using JWST NIRSpec-PRISM and MIRI LRS, respectively. The spectra are considered to be ideal, and do not account for the effects of stellar activity or variability. In the transit spectrum, it can be seen that the sulfuric acid (H$_2$SO$_4$) haze greatly truncates the spectrum, reducing the size of the CO$_2$ features and completely concealing the lone SO$_2$ feature \citep{ostberg2023c}. The presence of haze will make it difficult to detect CO$_2$, and can also make it ambiguous to decipher whether the planet has a thin, CO$_2$ dominated atmosphere, or a thick, hazy atmosphere like Venus. Whereas some features are still visible in the hazy transit spectrum, the CO$_2$ features are completely concealed in the hazy eclipse spectrum. The eclipse spectrum of the same planet with no atmosphere is also shown for reference.

\begin{figure*}
    \begin{center}
        \begin{tabular}{cc}
            \includegraphics[width=8.5cm]{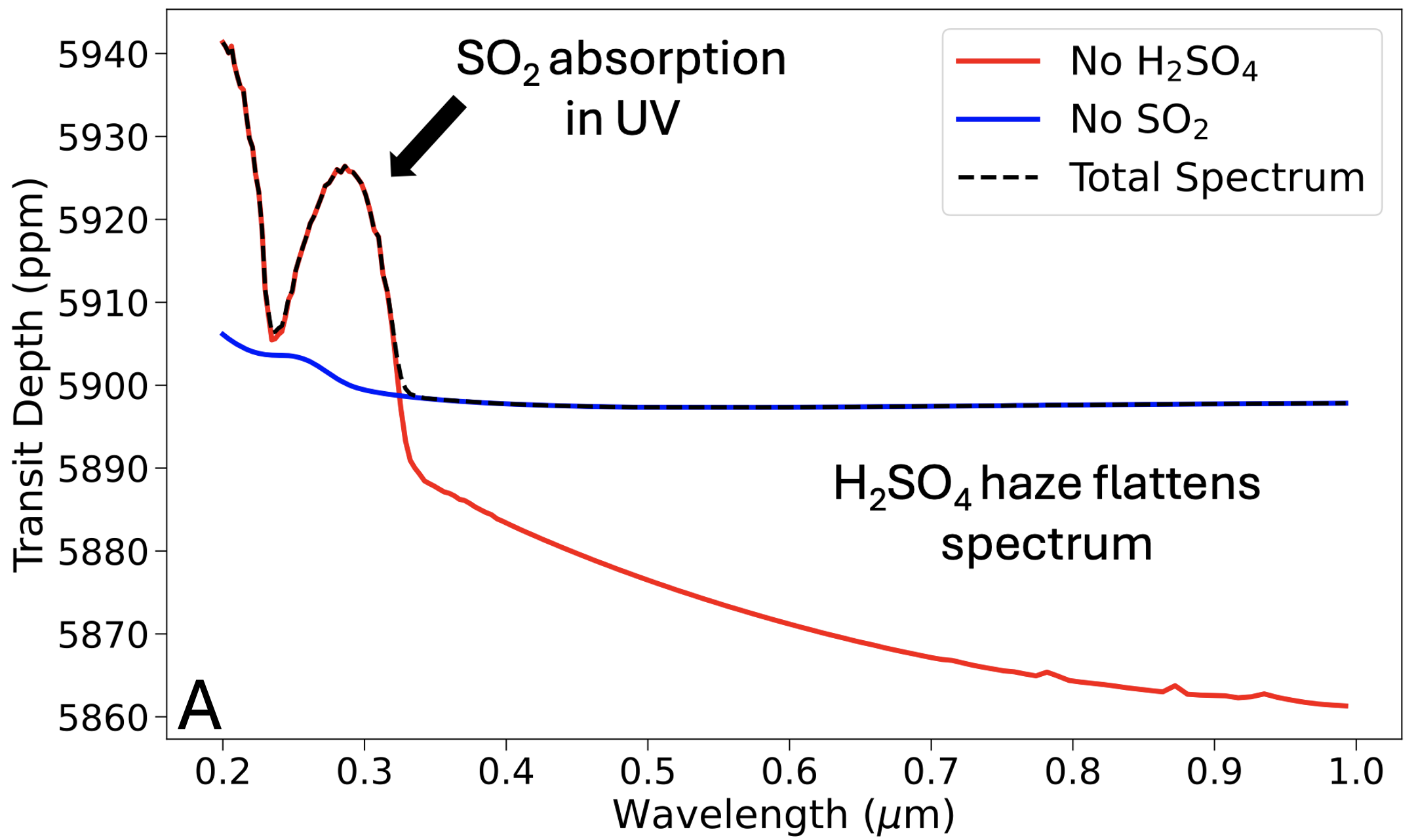} &
            \includegraphics[width=8.5cm]{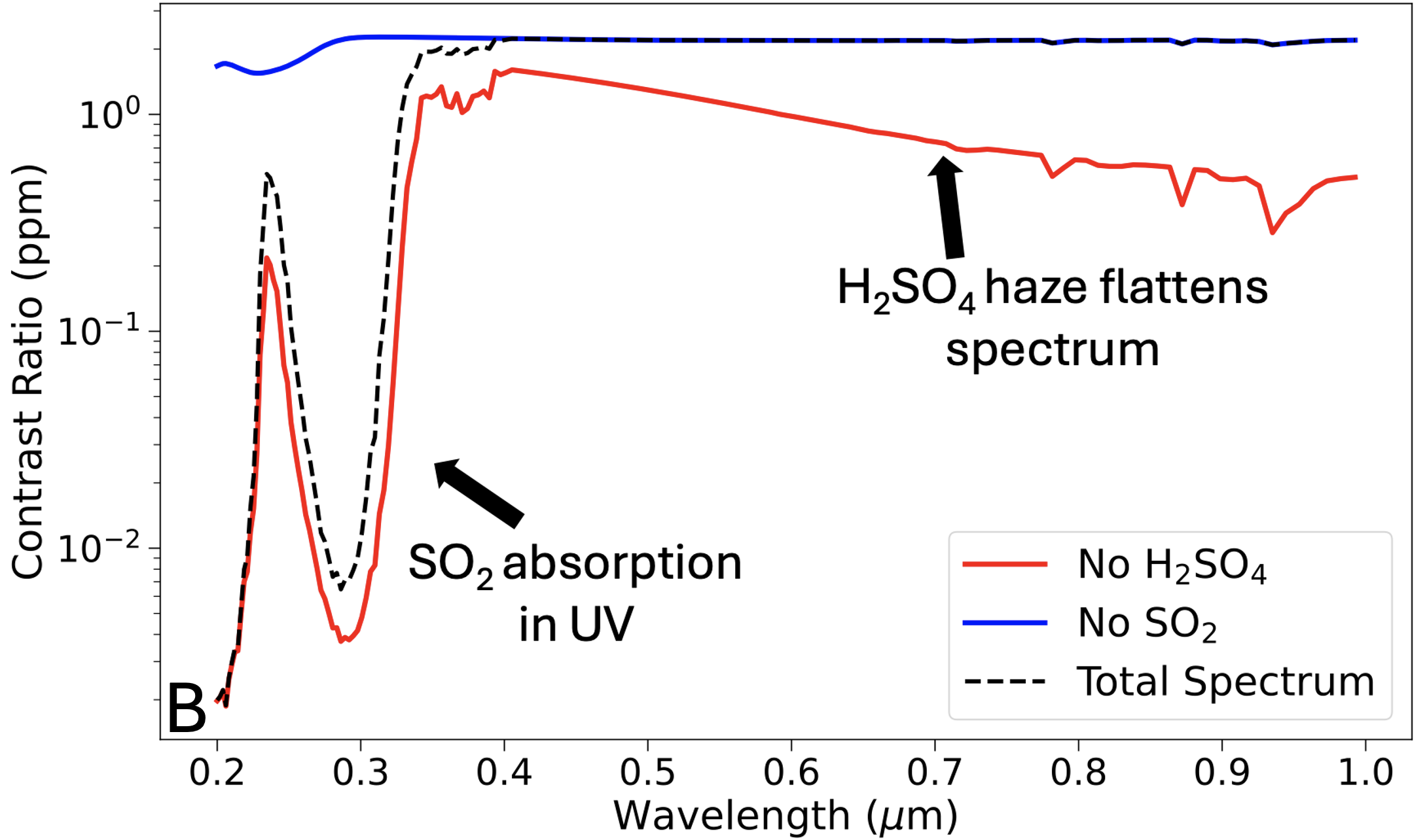} \\
            \includegraphics[width=8.5cm]{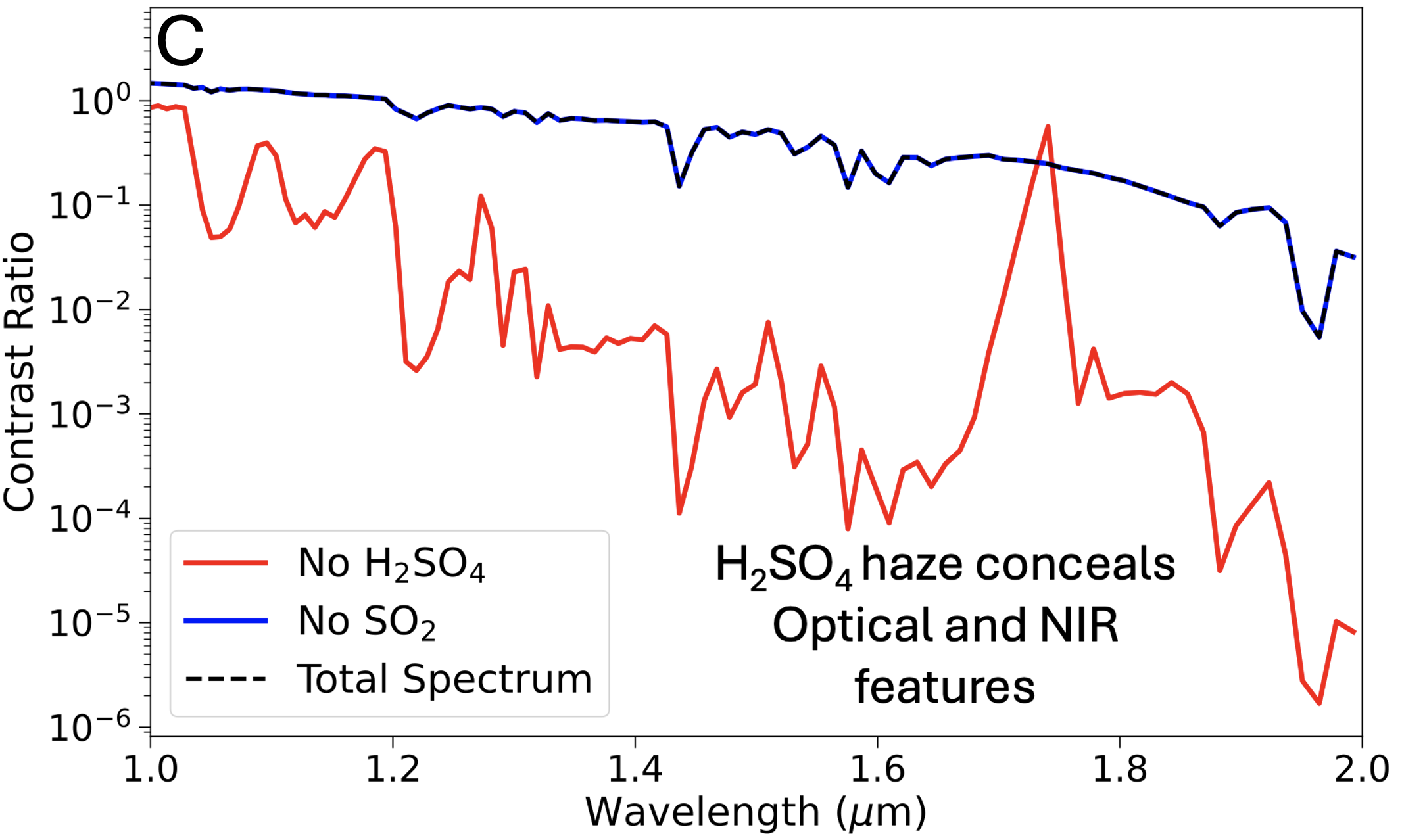} &
            \includegraphics[width=8.5cm]{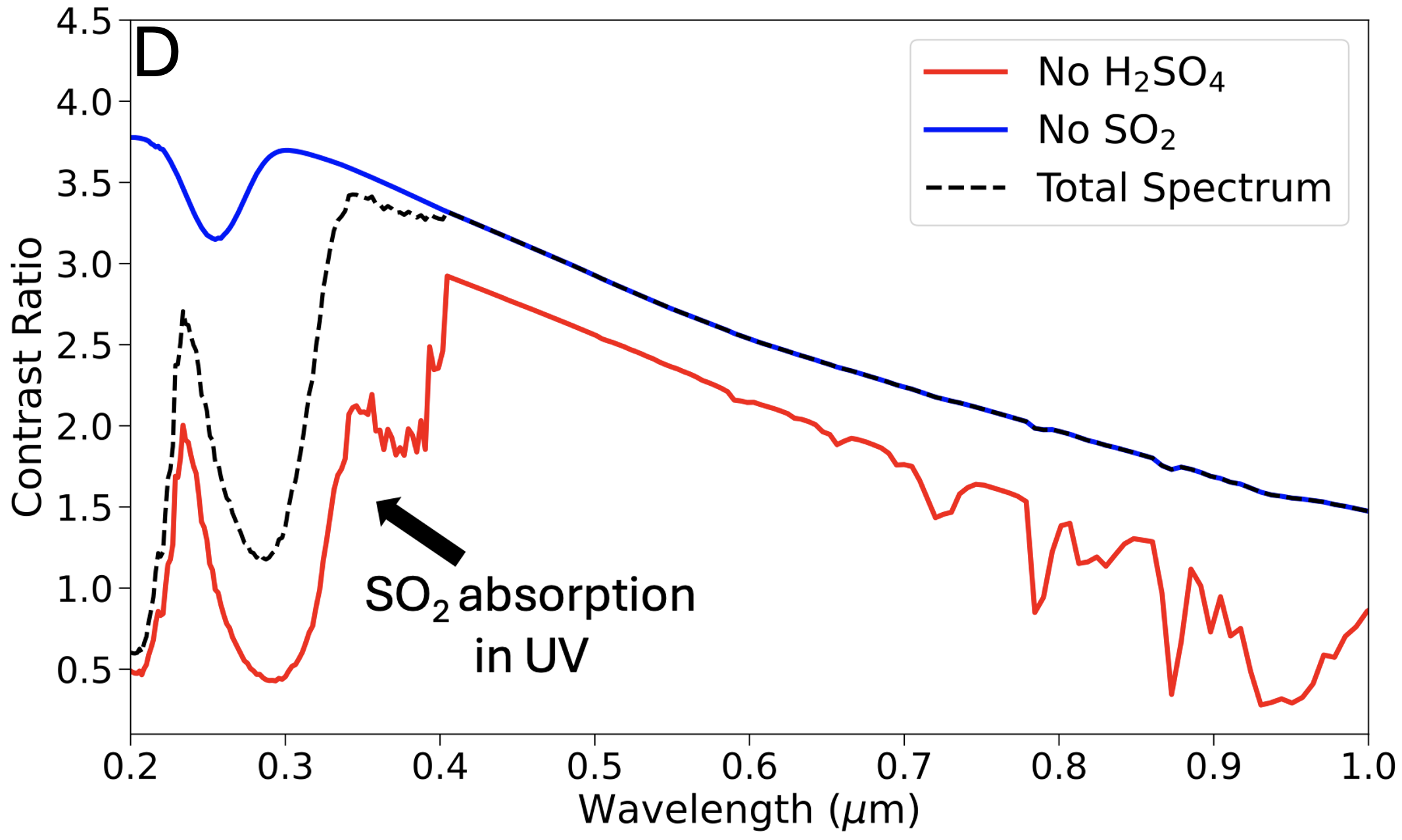} \\
        \end{tabular}
    \end{center}
  \caption{Simulated HWO spectra of an exoplanet with a 10-bar, Venus-like atmosphere. The upper panels assume the planet is orbiting an M-type star, whereas the lower panels assume a solar-type star. Top-left: UV and optical transmission spectrum, showing SO$_2$ absorption. Top-right: UV and optical reflectance spectrum during secondary eclipse. Bottom-left: Reflectance spectrum at optical and NIR wavelength from direct imaging observations with HWO. Bottom-right: Reflectance spectrum at UV and optical wavelengths from direct imaging observations with HWO.}
  \label{fig:spec}
\end{figure*}

To illustrate the benefits of the HWO bandpasses in comparison to JWST, we also simulated observations of the same planet using a hypothetical HWO telescope (Figure~\ref{fig:spec}). Panels A and B assume the planet is orbiting an M-type star for demonstrating transit and eclipse observations, whereas panels C and D assume a solar-type host star for direct imaging observations. In each panel, the dotted black line represents the true spectrum, whereas the red and blue lines are the same spectrum without the effects of H$_2$SO$_4$ haze and SO$_2$, respectively. Panel A of Figure~\ref{fig:spec} shows a simulated HWO transmission spectrum for a bandpass of 0.2--1.0~$\mu$m. Sulfur dioxide (SO$_2$) produces a prominent signature at UV wavelengths despite the presence of haze, whereas the rest of the spectrum is considerably impacted by the haze. Panel B shows a simulated reflectance spectrum of the same exoplanet from secondary eclipse observations with HWO. The presence of SO$_2$ causes a large dip in the UV that is visible with haze present, but the H$_2$SO$_4$ haze change the slope of the spectrum at visible wavelengths. Using HWO for transit and eclipse spectroscopy could enable the detection of SO$_2$ and H$_2$SO$_4$ haze on Venus-like worlds, which would otherwise be difficult to achieve with JWST alone. Furthermore, HWO will provide the ability to directly detect the CO$_2$ bands near 1.05, 1.21, and 1.42~$\mu$m, and the absence of significant H$_2$O detection at 0.94, 1.15, and 1.4~$\mu$m may rule out habitable scenarios and help resolve potential degeneracies between O$_3$ and SO$_2$ in the UV based \citep{loftus2019}.

Panel C of Figure~\ref{fig:spec} shows a simulated reflectance spectrum at optical and NIR wavelengths of the same exoplanet orbiting a solar-type star from HWO direct imaging observations. SO$_2$ has little effect on the spectrum at these wavelengths, however H$_2$SO$_4$ clouds conceal underlying SO$_2$ features. HWO may be able to identify hazy, Venus-like worlds by detecting similarly flattened reflectance spectra. Finally, the bottom-right panel of Figure~\ref{fig:spec} shows a simulated reflectance spectrum at UV and optical wavelengths of the same exoplanet from direct imaging observations with HWO. Although SO$_2$ absorption has been detected with JWST near 4.05~$\mu$m for a giant planet \citep{tsai2023b}, a substantial SO$_2$ absorption feature is also present in the UV that could be detected with HWO. Detecting SO$_2$ would provide crucial information for inferring whether an exoplanet is volcanically active, or ruling out that it has a wet, Earth-like atmosphere. The wavelengths accessible to HWO will thus make it a powerful resource for the study of Venus-like exoplanets, and will provide insights into the likelihood of Venus-like evolutions for terrestrial planets that cannot be achieved with JWST alone. Additionally, the potential for transit and eclipse spectroscopy with HWO should not be overlooked given the further science that such observations will enable. 


\subsection{Spectropolarimetry}

As previously discussed, the thick cloud deck of a Venus-like planet would obscure a majority of its atmosphere and flatten its spectra, making characterization strategies that only rely on its reflected or transmitted (i.e., unpolarized) flux difficult \citep[e.g.,][]{lustigyaeger2019b}. Instead, spectropolarimetry of the directly imaged light from exoVenuses could provide a more powerful method with which to identify and characterize their atmospheres and clouds \citep[e.g.,][]{mahapatra2023}. For example, by comparing observations to numerical simulations of polarized phase curves, \citet{hansen1974c} provided the first accurate description of the microphysics of the Venusian clouds, showing that the clouds were optically thick and contained spherical liquid droplets composed of a sulfuric acid solution with radii of $\approx1.05$~$\mu$m in a narrow size distribution. These cloud particle size and composition estimates were later confirmed by in-situ measurements performed by the \textit{Pioneer Venus} spacecraft \citep[e.g.,][]{knollenberg1980b,kawabata1980}, thereby highlighting the power of polarimetry to characterize planetary atmospheres in ways flux alone cannot.

\begin{figure*}
    \begin{center}
        \begin{tabular}{cc}
            \includegraphics[width=8.6cm]{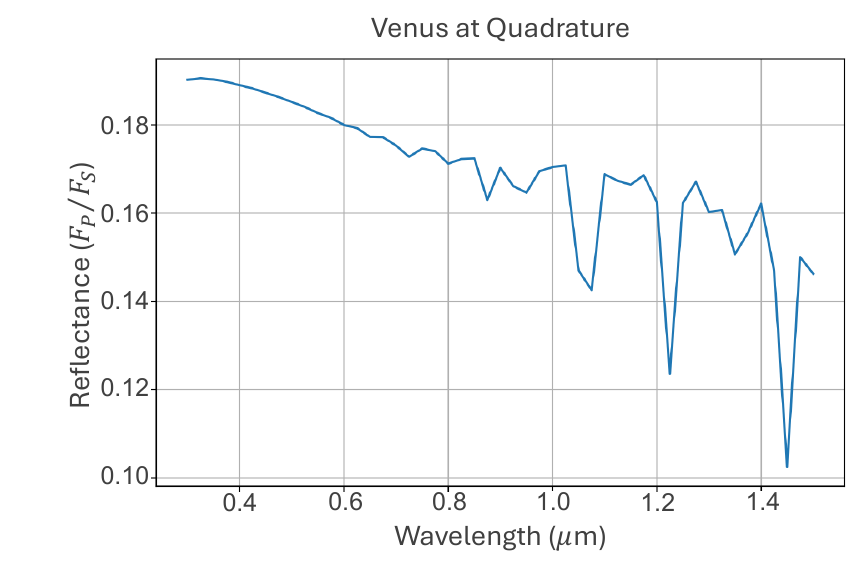} &
            \includegraphics[width=8.4cm]{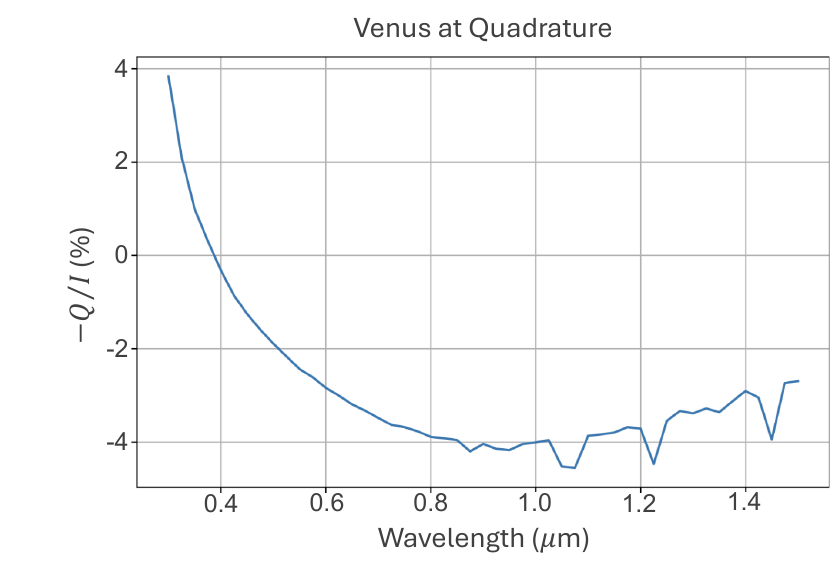} \\
        \end{tabular}
    \end{center}
  \caption{Simulated unpolarized reflectance (left) and signed degree of linear polarization (right) spectra for a Venus-like exoplanet around a solar-type star. The optically thick H$_2$SO$_4$ clouds flatten both spectra and suppress the NIR spectral features in the polarization, while the CO$_2$ Rayleigh scattering strongly polarizes the light at the shorter wavelengths.}
  \label{fig:polar}
\end{figure*}

An example of the level of polarization from exoVenuses that we expect to achieve with HWO can be seen in Figure~\ref{fig:polar}. Here we simulated a Venus-like exoplanet at quadrature around a solar-type star with the Versatile Software for Transfer of Atmospheric Radiation (VSTAR) polarization-enabled radiative transfer code \citep[][]{bailey2018c}, which utilizes the discrete-ordinate method from the VLIDORT \citep[e.g.,][]{kopparla2018} polarized light solution \citep[][]{spurr2006}. The model atmosphere was vertically heterogeneous, split into 50 horizontally homogeneous and plane parallel layers, and used the composition and temperature-pressure profiles from the Venus International Reference Atmosphere \citep[VIRA;][]{kliore1985b,moroz1997}. The vertical optical thickness profile for the clouds was based on those of the four particle modes described in \citet{crisp1986a}. The left panel of Figure~\ref{fig:polar} shows the resulting visible to near-infrared (VNIR) unpolarized reflectance spectrum, and the right panel shows the VNIR spectrum for the signed degree of linear polarization, $-Q / I$. The signed degree of linear polarization holds information about the direction of the polarized light, and as it ranges from -100\% to 100\%, larger deviations from 0 in either the positive or negative direction indicate higher levels of polarization \citep[see, e.g.,][for more details]{karalidi2011}. Similarly to Figures~\ref{fig:jwstspectra} and \ref{fig:spec}, the opaque H$_2$SO$_4$ clouds flattens both spectra and suppresses the NIR absorption features, especially in polarization. However, the large abundance of CO$_2$ in the atmosphere creates high levels of Rayleigh scattering that strongly polarize the light in the UV, leading to a steep blue slope. As this Rayleigh slope is not seen in the unpolarized reflectance, utilizing polarized light measurements could therefore assist with the detection of CO$_2$ in the atmosphere of a Venus-like planet.

The spectra in Figure~\ref{fig:polar} were generated only for the planet at a phase angle ($\alpha$) of $90\degr$, as this phase will provide the largest angular separation between the planet and its host star to optimize direct imaging observations. However, measuring spectra at only one $\alpha$ can limit planetary characterization, especially since studies have shown that Venus exhibits some of its lowest levels of polarization around quadrature \citep[e.g.,][]{hansen1974c}. Instead, observing a planet across multiple points in its orbit can allow for more accurate detections and characterizations of atmospheric and surface features \citep[e.g.,][]{vaughan2023}. Polarized phase curves are especially powerful for characterizing planetary atmospheres, as the locations and numbers of different features in the curves can be used to determine the compositions and optical properties of clouds and hazes \citep[e.g.,][]{goodisgordon2025a,goodisgordon2025b}. Polarization could assist with discerning whether a rocky exoplanet possesses H$_2$O clouds like Earth, which produce a rainbow peak at $\alpha \approx 40\degr$, or H$_2$SO$_4$ clouds like Venus, which produce a ``rainbow" peak at $\alpha \approx 20\degr$. It will therefore be vitally important to consider spectropolarimetry in future observations of terrestrial exoplanets.


\subsection{Synergy with Venus Missions}

There are several planned upcoming missions to Venus, including the NASA VERITAS (Venus Emissivity, Radio Science, InSAR, Topography, and Spectroscopy) \citep{cascioli2021} and DAVINCI (Deep Atmosphere Venus Investigation of Noble gases, Chemistry, and Imaging) \citep{garvin2022} missions, the ESA EnVision spacecraft \citep{widemann2023}, and the privately funded Morning Star missions \citep{seager2022a}. In terms of polarimetric observations, neither VERITAS nor DAVINCI will possess polarimeters. However, EnVision will carry the VenSpec-H spectrometer, which will contain two linear polarization filters covering two near-infrared spectral bands that can collect data on the microscopic properties of the Venusian clouds and hazes \citep[e.g.,][]{widemann2023}. On the other hand, spectroscopic follow-up observations of exoVenus targets will face additional challenges, such as clouds/hazes producing atmospheric opaqueness \citep{lustigyaeger2019b}, potentially obscuring spectral features that would typically be used to distinguish between Venus-based and Earth-based atmospheres \citep{ehrenreich2012a,garciamunoz2012b,barstow2016a,lincowski2018,bergin2023,ostberg2023b,calder2025}. The in-situ and orbital results from the coming Venus missions will provide significant improvements for cloud microphysics, sulfur chemistry, surface-atmosphere exchange, and isotopic constraints, which will propagate into exoplanet retrieval priors and interpretations. Thus, this tight Solar System to extrasolar feedback loop will improve the fidelity of exoVenus atmospheric inferences and at the same time will return critical information to Venus climate studies \citep{kane2024b}.


\section{Discussion}
\label{discussion}

A central interpretive challenge for exoVenus characterization is the triad of degeneracies among aerosol properties, gas abundances, and temperature structure. In reflected light, H$_2$SO$_4$ clouds can mute narrow features and impose a red/gray continuum slope that mimics low molecular absorption; conversely, UV SO$_2$ bands can be masked if clouds extend to high altitudes. Joint UV+optical/NIR coverage, ideally contemporaneous, therefore targets orthogonal parts of the parameter space: UV constrains SO$_2$ and O$_3$ column abundances and photochemical regimes, and optical/NIR reflectance fixes cloud optical depth and particle sizes via continuum shape and weak bandheads, and polarimetry isolates droplet composition through the phase angle and amplitude of the rainbow feature \citep{hansen1974c,bailey2007b,arney2018,lincowski2018,goodisgordon2025a,woitke2025,roccetti2025b}. Retrieval experiments show that even modest S/N UV points can collapse large volumes of aerosol/gas covariance when paired with optical–NIR spectra of comparable S/N and spectral resolution \citep{villanueva2018b,meadows2018c}. HWO's ability to gather these measurements for the same target within a single observing campaign is thus not just a convenience but a requirement to achieve robust inferences.

The Venus cloud deck exhibits a near-UV/blue absorber that depresses the dayside albedo over 0.32--0.50~$\mu$m and is constrained to the upper cloud region ($\gtrsim 57$~km), consistent with absorption embedded within cloud particles \citep{shimizu1977,toon1982,perezhoyos2018,dai2025}. Importantly, the absorber's UV behavior overlaps classical SO$_2$ absorption ($\lambda \lesssim 320$~nm), and disk-integrated Akatsuki analyses demonstrate that SO$_2$ and the unknown absorber can be spectrally degenerate at 283~nm \citep{lee2021b}. Laboratory and modeling work further suggest that multiple chemically distinct candidates can match the broadband UV–blue features, reinforcing that low-/moderate-resolution UV–visible spectra may not uniquely identify the absorber and must instead marginalize over it \citep{jiang2024c}. For HWO, there are several reasons why this uncertainty may directly impact near-UV and visible reflected-light characterization. First, because a Venus-like absorber would reduce planet–star contrast in the bandpass used to constrain UV diagnostics. Second, it would introduce strong degeneracies between cloud-top pressure/optical depth and UV absorbers (SO$_2$, O$_3$, and the unknown aerosol absorber). Third, the uncertainty would propagate into systematic abundance uncertainties when spectra are binned and continuum calibration is limited by wavelength-dependent starlight-suppression residuals \citep{tuchow2024,tuchow2025a}. These effects are potentially amplified for exoVenus analogs because thick H$_2$SO$_4$ clouds already mute gaseous features, making the unknown UV absorber a potential source of confusion that can bias classification and interpretation of HWO UV–visible spectra \citep{kane2019d,way2023a}.

Stellar and system context further complicate, and enrich, interpretation. Stellar activity modulates UV irradiance, driving time-dependent photochemistry; activity diagnostics from EPRV and contemporaneous UV monitoring can be assimilated into retrieval priors to avoid biasing SO$_2$/O$_3$ inferences \citep{fischer2016,morgan2021b}. Host spectral type matters twice: it sets both the planet’s bolometric flux and the UV/blue photon supply that governs sulfur oxidation pathways and haze formation, with K dwarfs offering advantageous contrasts and duty cycles for direct imaging. Stellar UV photons regulate the upper-atmosphere chemical state of Venus-like planets and therefore modify their reflected-light spectra in ways that directly impact detectability. In CO$_2$-dominated Venus-like atmospheres, CO$_2$ photolysis and UV-driven sulfur oxidation control whether SO$_2$ is removed above the clouds and converted into H$_2$SO$_4$ cloud material \citep[e.g.,][]{yung1982b,hu2013}. For a Venus twin irradiated by a G-star spectrum, solar UV efficiently photochemically depletes SO$_2$ in the observable (above-cloud/haze) column and drives H$_2$SO$_4$ cloud formation, implying that SO$_2$ absorption in near-UV reflectance can be intrinsically weak or absent even for a truly dry, Venus-like surface state \citep{jordan2025a}. In contrast, photochemical models for cooler K--M hosts show orders-of-magnitude reductions in SO$_2$/OCS/H$_2$S depletion above the clouds, and for quiescent M-dwarf spectra the H$_2$SO$_4$ cloud-forming inventory can drop substantially, potentially reducing cloud opacity while simultaneously deepening sulfur-gas UV absorption features in reflected light \citep{jordan2021}. Because M-dwarf UV spectra are line-dominated and time-variable, robust interpretation of any SO$_2$/O$_3$-related absorption in the HWO near-UV bands requires stellar-UV-informed priors and, where possible, contemporaneous UV activity monitoring \citep{france2013a,damiano2023b,morgan2024b}. As previously described, system architecture (e.g., exterior gas giants) may regulate volatile delivery, mantle oxidation state, and long-term climate stability \citep{raymond2014d,raymond2017b,venturini2020b,vervoort2022,kane2024a,kane2025a}. Demographic analyses that depend on architecture are therefore essential for causal inference of terrestrial planet evolution \citep{kopparapu2014,kane2024b}. On the planetary side, mass and radius (from RVs/TTVs) determine gravity and scale heights that feed directly into the detectability of UV features, especially in high-mean-molecular-weight atmospheres. These measurements also inform interior models that bound volatile budgets and surface pressures, closing the inference loop from spectra to climate state \citep{kane2014e,way2020}.

An additional theme is the safeguarding of biosignature science for temperate targets by treating exoVenuses as intentionally ``non-biosignature'' calibrators. The same false-positive oxygen pathways discussed for HZ planets (e.g., hydrogen escape following massive photolysis) operate more efficiently at VZ insolations, producing comparable oxygen signatures to those expected from biotic sources \citep{domagalgoldman2014,wordsworth2014}. Thus, quantifying these false-positive biosignatures on hot, cloudy worlds improves the priors used to interpret any future oxygen detections in the HZ \citep{arney2016,meadows2017,meadows2018c}. Conversely, credible SO$_2$ detections on sub-$2R_\oplus$ planets would implicate ongoing volcanism or sustained sulfur cycling, with implications for interior thermal states and resurfacing frequencies; combining UV constraints with time-variable optical polarization (cloud patchiness) could, in principle, separate episodic from steady-state outgassing. Finally, lessons from Venus missions (vertical cloud structure, microphysical pathways to H$_2$SO$_4$, noble-gas and isotopic constraints on volatile loss) translate directly into retrieval parameterizations and priors that reduce model non-uniqueness \citep{garvin2022,cascioli2021,widemann2023,kane2019d,kane2021d}.

From an operations perspective, three design/strategy implications recur. First, UV access to at least 0.3~$\mu$m (preferably to 0.2~$\mu$m) is a high scientific multiplier for exoVenus science because it supplies unique sulfur/redox leverage not available to JWST and breaks aerosol/gas degeneracies in reflected light. Second, small inner working angles (IWA) are needed to routinely access 0.3--1.0~AU for nearby FGK stars, which is where many VZ planets reside (see Section~\ref{sample}); this requirement couples to coronagraph/starshade performance and drives target completeness. Third, phase-resolved spectroscopy and polarimetry at selected phase angles (bracketing the rainbow) should be incorporated into scheduling to retrieve droplet composition and cloud vertical structure, not treated as follow-on observations \citep{hansen1974c,bailey2007b,goodisgordon2025a,roccetti2025b}. These considerations directly inform exposure-time calculators, yield simulations, and survey designs that will be needed to translate science priorities into observing programs \citep{villanueva2018b,kane2024e}.


\section{Conclusions}
\label{conclusions}

HWO offers a path to transform our understanding of Venus-like climates by (i) measuring atmospheric sulfur chemistry and aerosol microphysics through joint UV+optical/NIR spectroscopy and spectropolarimetry; (ii) assembling a statistically meaningful sample of sub-$2R_\oplus$ planets at VZ fluxes across FGK hosts; and (iii) closing inference loops with masses, radii, stellar abundances, and Solar System ground truth. The scientific returns of pursuing observations of exoVenus candidates include the robust identification of post-runaway greenhouse states and quantitative constraints on their occurrence as functions of stellar type and architecture, thus placing Venus within a much broader context of terrestrial planet evolution. The longer-term dividends extend to biosignature science by establishing abiotic oxygen baselines and by stress-testing retrieval frameworks under the most challenging aerosol regimes.

Venus and exoVenus science is therefore an important design driver for HWO capabilities: wavelength coverage from 0.2--$\gtrsim$1.5~$\mu$m with sufficient spectral resolution to separate continua from bandheads; polarimetric sensitivity and calibrated phase control to capture rainbow features; and coronagraphic/starshade performance yielding IWAs that regularly access 0.3--1.5~AU around the nearest FGK stars. With these capabilities, a survey of $\sim$30 nearby systems, informed by precursor EPRV and transit programs, can deliver the demographic and physical constraints required to place Earth and Venus in their proper galactic context \citep{kane2014e,kopparapu2014,way2020,kane2024b}. Although the specific design features of HWO are still a matter of considerable discussion, including the inclusion of a polarimeter, a mission structure that overlaps with our describe requirements would be of enormous benefit to the Venus community, and the exoplanet field as a whole. Furthermore, HWO synergy with the DAVINCI, VERITAS, EnVision, and Morning Star missions will sharpen priors, reduce degeneracies, and ensure that inferences about worlds around other stars remain anchored to the only two terrestrial climates we can interrogate directly \citep{kane2019d,cascioli2021,kane2021d,garvin2022,widemann2023,seager2022a}.


\section*{Acknowledgements}

S.R.K acknowledges support from the NASA Astrophysics Decadal Survey Precursor Science (ADSPS) program under grant No. 80NSSC23K1476. K.B acknowledges access to the Klone Hyak Supercomputer through the Virtual Planetary Laboratory. The results reported herein benefited from collaborations and/or information exchange within NASA's Nexus for Exoplanet System Science (NExSS) research coordination network sponsored by NASA's Science Mission Directorate.


\software{Planetary Spectrum Generator \citep[PSG;][]{villanueva2018b}, Versatile Software for Transfer of Atmospheric Radiation \citep[VSTAR;][]{bailey2018c}}




\end{document}